\documentclass[aps,prl,twocolumn,showpacs,preprintnumbers,amsmath,amssymb,superscriptaddress]{revtex4}
\usepackage{graphicx}
\usepackage{dcolumn}
\usepackage{bm}
\begin{document}
\title{Measurements of ultracold neutron lifetimes in solid deuterium}

\author{C. L. Morris}\email[electronic address:]{CMorris@lanl.gov}\affiliation{Los Alamos National Laboratory, Los Alamos, NM 87544, USA}

\author{J. M. Anaya}\affiliation{Los Alamos National Laboratory, Los Alamos, NM 87544, USA}

\author{T. J. Bowles}\affiliation{Los Alamos National Laboratory, Los Alamos, NM 87544, USA}

\author{B. W. Filippone}\affiliation{Kellog Radiation Laboratory, California Institute of Technology, Pasandena, Ca 91125, USA}

\author{P. Geltenbort}\affiliation{Institut Laue-Langevin, BP 156, F-3804 Grenoble cedex 9, France}

\author{R. E. Hill}\affiliation{Los Alamos National Laboratory, Los Alamos, NM 87544, USA}

\author{M. Hino}\affiliation{University of Kyoto, Kyoto 606-8501, Japan}

\author{S. Hoedl}\affiliation{Princeton University, Princeton, NJ 08544, USA}

\author{G. E. Hogan}\affiliation{Los Alamos National Laboratory, Los Alamos, NM 87544, USA}

\author{T. M. Ito}\affiliation{Kellog Radiation Laboratory, California Institute of Technology, Pasandena, Ca 91125, USA}

\author{T. Kawai}\affiliation{University of Kyoto, Kyoto 606-8501, Japan}

\author{K. Kirch}\altaffiliation[Present address:] {Paul Scherrer Institut,Switzerland}\affiliation{Los Alamos National Laboratory, Los Alamos, NM 87544, USA}

\author{S. K. Lamoreaux}\affiliation{Los Alamos National Laboratory, Los Alamos, NM 87544, USA}

\author{C.-Y. Liu}\affiliation{Princeton University, Princeton, NJ 08544, USA}

\author{M. Makela}\affiliation{Virginia Polytechnical Institute and State University, Blacksburg, Va 24061, USA}

\author{L. J. Marek}\affiliation{Los Alamos National Laboratory, Los Alamos, NM 87544, USA}

\author{J. W. Martin}\affiliation{Kellog Radiation Laboratory, California Institute of Technology, Pasandena, Ca 91125, USA}

\author{R. N. Mortensen}\affiliation{Los Alamos National Laboratory, Los Alamos, NM 87544, USA}

\author{A. Pichlmaier}\affiliation{Los Alamos National Laboratory, Los Alamos, NM 87544, USA}

\author{A. Saunders}\affiliation{Los Alamos National Laboratory, Los Alamos, NM 87544, USA}

\author{S. J. Seestrom}\affiliation{Los Alamos National Laboratory, Los Alamos, NM 87544, USA}

\author{D. Smith}\altaffiliation[Present address:] {Stanford Linear Accelerator Center, USA}\affiliation{Princeton University, Princeton, NJ 08544, USA}

\author{W. Teasdale}\affiliation{Los Alamos National Laboratory, Los Alamos, NM 87544, USA}

\author{B. Tipton}\affiliation{Kellog Radiation Laboratory, California Institute of Technology, Pasandena, Ca 91125, USA}

\author{M. Utsuro}\altaffiliation[Present address:] {Research Center for Nuclear Physics, Osaka University, Japan}\affiliation{University of Kyoto, Kyoto 606-8501, Japan}
   
\author{A. R. Young}\affiliation{North Carolina State University, Raleigh, NC 27695, USA}

\author{J. Yuan}\affiliation{Kellog Radiation Laboratory, California Institute of Technology, Pasandena, Ca 91125, USA}

\date{\today}

\begin{abstract}
We present the first measurements of the
survival time of ultracold neutrons (UCNs) in solid deuterium (SD$_2$). 
This critical
parameter provides a fundamental limitation to the effectiveness of
superthermal UCN sources that utilize solid ortho-deuterium
as the source material. 
Superthermal UCN sources offer orders of magnitude
improvement in the available densities of UCNs, and are
of great importance to fundamental particle-physics experiments
such as searches for a static electric dipole moment and lifetime
measurements of the free neutron.
These measurements are performed utilizing a
SD$_2$ source coupled to a spallation source of neutrons, providing
a demonstration of UCN production in this geometry and
permitting systematic studies of the influence of thermal up-scatter and
contamination with para-deuterium on the UCN survival time.
\end{abstract}

\maketitle

Neutrons with kinetic
energies less than 340 neV (corresponding to a temperature T$<$ 5mK) can
be trapped in material bottles and are referred to as
ultracold neutrons (UCNs) \cite{Zel59,Lus69,Ste69}. 
UCN densities at reactor sources have gradually increased
with reactor power and improved techniques for extracting the UCN flux. The
highest bottled densities reported in the literature, 41/cm$^3$, have been obtained at the Institut Laue-Langevin (ILL)
reactor in Grenoble \cite{Ste86}.

Measurements of the neutron electric dipole moment \cite{Har99,Alt96}
and the neutron lifetime \cite{Mam93,Arz00,Pic00} attest to the
utility of bottled UCNs for fundamental experiments with neutrons. UCNs may
prove useful in improved measurements of angular correlations in neutron
beta-decay \cite{Aproposal,Gar01}, although experiments of this kind utilizing UCNs have not yet
been performed. All of these experimental programs have been limited by the
available densities of UCNs. 

Superthermal UCN production, where the production rate
of UCNs due to down-scattering in energy is larger than the combined up-scatter and nuclear-absorption rates
in the material, was first proposed in 1975 by Golub and
Pendlebury \cite{Gol75} in superfluid $^{4}$He and experimentally investigated shortly thereafter 
\cite{Alt80,Gol83}. In this process phonon creation in the liquid is used to 
down-scatter cold neutrons to the UCN regime, while up-scattering 
is suppressed by maintaining the superfluid at sufficiently low temperature.
Because $^{4}$He has no nuclear absorption, the only limitations to the density of UCNs
accumulated are wall losses and neutron beta decay.  The production
of UCNs by this process has been observed and agrees with 
theoretical expectations \cite{Gol83,Yo92,Kil87,Huf00,Bro01}. 

While superfluid $^{4}$He is an excellent superthermal converter, a few other materials, such as
solid deuterium (SD$_2$), satisfy 
the criteria for superthermal production.
The limiting UCN density, $\rho _{UCN}$, one can obtain using a SD$_2$
source is given by the product of the rate of UCNs
production in the solid, $R$, and the lifetime of UCNs in the solid, $\tau
_{SD}$: $\rho _{UCN}=R\tau _{SD}$. A storage bottle opened to such a
source will come into density equilibrium with the density in the solid.
This led to the proposal of a thin-film source where the inside of a neutron
bottle is coated with a thin layer of SD$_2$ and the bottle is embedded in a
cold neutron flux \cite{Gol83a,Yu86}. The volume comes into equilibrium with
the UCN density with a time constant for the coupled system,$\tau $ , given
by, 
\begin{equation}
\tau =\tau _{SD}\frac{V}{V_{SD}},  \label{eqn:tau}
\end{equation}
where $V_{SD}$ is the volume of SD$_2$ in the source and $V$ is the
total volume of the storage bottle. A valuable summary of SD$_2$
thin-film sources is presented in \cite{Gol91}. The effects of gravity and of
the potential of the solid as well as UCN losses other than absorption in
the film have been neglected in this expression but do not fundamentally alter the above
picture. Ultimately, the limit to the UCN density
is established by the trade-off between the cold-neutron flux intensity and
energy distribution (which determine the production rate) and the heat
deposited by neutrons and gamma rays in the SD$_2$ and bottle walls,
because $\tau _{SD}$ is a strong function of temperature. However, the
predicted production rates in SD$_2$ \cite{Yu86} and lifetimes \cite{Liu00} have not yet been quantitatively verified.  Efforts to utilize
SD$_2$ sources at reactors identified possible gains but suffered from 
problems with cooling the solid at full reactor power \cite{Alt80}.

Pokotilovski pointed out the advantages of UCN production in SD$_2$
at pulsed neutron sources \cite{Pok95} and showed the UCN densities 4--5 orders 
of magnitude greater than existing reactor-based UCN sources might be possible.
More recently, 
the use of spallation as a pulsed source has been suggested \cite{Ser97,Ser00}. 
In a spallation UCN source, cold neutrons are produced 
by moderating spallation neutrons produced in a
heavy target by a medium-energy pulsed proton beam. 
These cold neutrons are used to drive a SD$_2$ superthermal converter. 
In a spallation target,
the amount of heating for each neutron is lower than in a reactor, allowing
higher neutron densities in the vicinity of a spallation target to be achieved. 
Even higher neutron densities can be ontained 
by pulsing the proton beam and valving
off the UCN storage volume from the production volume when the beam is off and
by using the time
when the beam is off to remove heat from the deuterium. In this case, the
maximum UCN density that is produced is limited only by the impulse heating
of the SD$_2$. Experiments with the stored UCNs can be performed
while the beam is off, eliminating the backgrounds due to capture gammas
found near continuously operated reactor sources.

In order to test the concept of a SD$_2$-based spallation UCN source,
we have built a test source that we have operated with single pulses of
protons produced by the LANSCE 800 MeV proton accelerator at Los Alamos
National Laboratory. In this letter we report the first measurements of $%
\tau _{SD}$, the lifetime of UCNs in SD$_2$. Our measurements
clearly demonstrate the critical influences of heating and para-deuterium
contamination on the UCN lifetime, and provide a quantitative foundation for
the development of SD$_2$ superthermal sources.

\begin{figure}
\includegraphics [width=3in,height=2.25in]{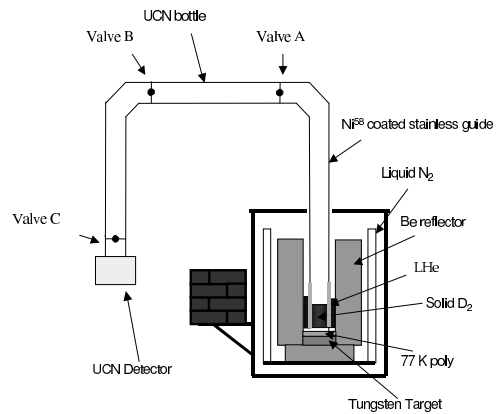}
\caption{\label{Schematic} Schematic view of the apparatus used for this experiment. Neutrons
were bottled in the region between valves A and B. Valve C was used to
insert a thin $^{58}$Ni-coated aluminum foil in the guide in front of the
detector. SD$_2$ measurements were made by counting the number of
UCNs that survive in contact with the deuterium as a function of time using
valve A, with valve B and C opened.}
\end{figure}

A schematic view of our apparatus is shown in Fig. \ref{Schematic}.
Spallation neutrons were produced in a tungsten target with short (typically
less than 160 ns long) pulses of 800 MeV protons from the LANSCE accelerator. 
The fast neutron flux was contained and amplified using
(n,2n) reactions in a layer of beryllium surrounding the spallation target.
The spallation neutrons were moderated and cooled in a thin layer of
polyethylene surrounding a $^{58}$Ni-coated stainless steel guide tube with
an inner diameter of 7.8 cm. 
The polyethylene and the bottom of
the guide were cooled with liquid helium, and a layer of deuterium was
frozen on the inside of the guide. 
UCNs produced in the SD$_2$ were
confined by the guide tube and could be directed through a series of valves
to the UCN detector. 
Neutrons were detected in a 5-cm-thick multi-wire chamber detector filled
with a mixture of $^{3}$He at 5 mbar and CF$_4$ at 1 bar. The
low $^{3}$He pressure and the large bend angle in the guide resulted in a
high degree of selectivity for detecting UCNs in the apparatus. 
Data were acquired using a multi-scalar that was
started by the proton beam passing through a toroidal
pick-up coil and that scaled the count rate from the UCN detector.

Several effects limit the lifetime of UCNs in SD$_2$: up-scatter
from phonons in the solid \cite{Yu86}, up-scatter from para-deuterium
molecules in the solid \cite{Liu00}, absorption on deuterium, and
absorption on hydrogen impurities in the solid. Model calculations exist for
the contribution of all of these effects on the UCN lifetime in the solid.
The total loss rate is a sum of contributions from each of
the sources listed above: 
\begin{equation}
1/\tau _{SD}=1/\tau _{phonon}+1/\tau_{para}+1/\tau _{Dabs}+1/\tau _{Habs},
\end{equation}
with the loss rate due to phonon
up-scatter having different contributions from the ortho- and para-deuterium
in the solid. Establishing the experimental basis to validate these models
as both accurate and complete is essential for the design of a UCN source
based on the superthermal production mechanism in solid ortho-deuterium.

SD$_2$ was frozen in the lower part of the cryostat using a 
helium transfer refrigerator. The temperatures of the lower guide walls and of the
liquid-helium cryostat were monitored with an array of silicon diodes
mounted on the guide and the aluminum cryostat walls. The measured temperatures
tracked the vapor pressure curve of SD$_2$ well at
higher temperatures. The temperature of the solid at lower temperatures was
obtained by averaging the temperatures of two diodes mounted on the outside
of the guide wall. Later measurements made with diodes embedded in the
solid indicate these measurements are accurate to 1 K.

Both the hydrogen contamination and the para-fraction in the SD$_2$
were measured by means of rotational Raman spectroscopy on a gaseous sample
taken by warming the deuterium after the measurement \cite{Liu00}. 
These
measurements yielded values for the HD concentrations in the gas that
varied from 0.2--0.3\% with an uncertainty of about 0.1\%. 
Other
contamination was removed by a palladium membrane in the D$_{2}$ gas
system, before introducing D$_{2}$ to the cryostat. The para-fraction was
controlled by converting the D$_2$ to a near thermal equilibrium
otho/para ratio in an iron-hydroxide-filled cell \cite{Wei56} cooled to a temperature at
or slightly below the triple point. In this way the para-fraction was
reduced from room-temperature equilibrium value of 33\% to 2--4\%.
Intermediate values were obtained by mixing deuterium at room temperature
equilibrium with converted deuterium before freezing. The precision of the
para-fraction measurements was typically of order of 1\%.

The solid volume was measured by integrating the flow of gas while growing
the solid. The volume was checked when the solid was warmed and the gas was returned
to a buffer volume. The uncertainties in the
pressure dependence of the calibration of the flow meter (20\%), and
uncertainties in the guide volume and temperature combined to lead to an
uncertainty in the solid volume of about 20\%.

\begin{figure}
\includegraphics[width=3in,height=2.25in]{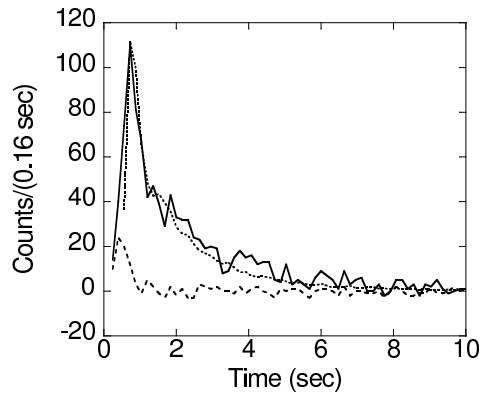}
\caption{\label{58NiBarrier}Background subtracted spectra with and without the $^{58}$Ni barrier in
the beam at the location of valve C. The dashed curve is the result of a
Monte Carlo calculation of UCN arrival times.}
\end{figure}

The sensitivity of the apparatus to UCNs was demonstrated
by measuring neutron arrival times with and without a 
$^{58}$Ni-coated 0.024-cm-thick aluminum foil in place at location C (in Fig.~\ref{Schematic}). 
These data are shown in Fig.~\ref
{58NiBarrier}. The number of counts, arriving in a time window between 0.5 sec and 10 sec, 
with the foil closed was 3\% of the number with the foil opened. 
About half of these could be
attributed to UCN leakage through the gap around the outside of the valve,
and the rest were neutrons with normal velocities sufficient to penetrate
the potential barrier provided by the $^{58}$Ni. 
These data unambiguously
demonstrate that the signal in the $^3$He detector at the end of the guide
was predominantly due to UCNs.

If gravity, wall losses, the SD$_2$ potential, and transport
effects are neglected, and if the SD$_2$ is thin enough so that its
volume is uniformly sampled by the neutrons, the lifetime of neutrons stored
in a bottle in contact with SD$_2$ is given by equation (\ref
{eqn:tau}). We have used this idea to measure the UCN lifetime in SD$_2$.
As depicted in Fig. \ref{Schematic}, valves B and C were open for
these measurements. UCNs were stored in the bottle between the end of the
guide and valve A, which includes the SD$_2$. The number of neutrons detected
in a 10 sec wide time gate after the time, $t$, when valve A was opened was measured, typically for $t=$0.5, 1., 2., and 4 seconds.
These data were fitted with the form $c_o
e^{-t/\tau}$. The parameters $c_o$ and $\tau$ were varied to 
produce the best fit for various conditions
of the SD$_2$. 
Background was subtracted by a fitting linear function to the data 
between 20 sec and 65 sec after the trigger and extrapolating under the data.
The $t_{SD}$ were extracted from the  $\tau$
using a lookup table of $t_{SD}$ \textit{vs} $\tau$ generated by fitting lifetime scans
modeled with Monte Carlo for different $t_{SD}$ that were analyzed in the same fashion
as the data. 


In the Monte Carlo transport, the full experimental geometry,
gravitation, the SD$_{2}$ potential (108 neV), and wall collisions in
the guide tube have been taken into account. 
There are a number of
parameters that must be determined from the experimental data in order to extract $t_{SD}$: the
probability of UCN loss and nonspecular reflection for each collision with
the guide wall, the SD$_2$ elastic scattering length,
and the physical configuration of the SD$_2$ frozen on the walls
(SD$_2$ can freeze as a flat ``pancake'' on the bottom of the
guide, or coat the entire He-cooled surface at the bottom of the guide, in a
``bucket''-shaped configuration).

Because we used very thin samples of SD$_2$ for most of the
lifetime measurements (0.4 cm), we were quite insensitive to
the scattering length for UCNs in SD$_2$. The saturation of UCN production
we observe for thicker samples (up to $\sim$6--cm of sample thickness) is consistent with
the large scattering length calculated by Hill \textit{et al.} \cite{Hill99}. 
However, our extracted lifetime results change by less than 10\% for scattering lengths as
small as 0.5 cm. We therefore used the theoretical value of 8 cm for the
elastic scattering length for all of the results presented here.

The ratio of diffuse to specular reflections for wall collisions was adjusted
to fit time-of-arrival spectra (see Fig. \ref{58NiBarrier}). The fraction of
wall losses per wall collision and the shape of the SD$_2$ ice were
adjusted in a combined fit to reproduce the volume dependence for all of the
low-temperature data. The extracted values for the guide parameters (0.025
for the ratio of diffuse to specular reflections) and wall losses 
(2.0$\times 10^{-3}$) were consistent with those extracted from an
independent set of measurements of guide transmissions and holding times at
the ILL. The lifetimes due to absorption on deuterium and
hydrogen were calculated using the known amount of hydrogen contamination,
the tabulated thermal cross sections and the $1/v$ dependence of the cross
sections. In addition the lifetimes in solid ortho- and para-deuterium due to temperature
up-scattering were taken from the literature \cite{Yu86,Liu00}. The
lifetime in solid para-deuterium due to molecular transitions was treated as
a free parameter and found to be $\tau _{SD}^{para}=1.2\pm 0.2$ ms, 
consistent with a calculation that gives {\bf $\tau
_{SD}^{para}=1.5$} ms \cite{Liu00}.

\begin{figure}
\includegraphics[width=3in,height=1.65in]{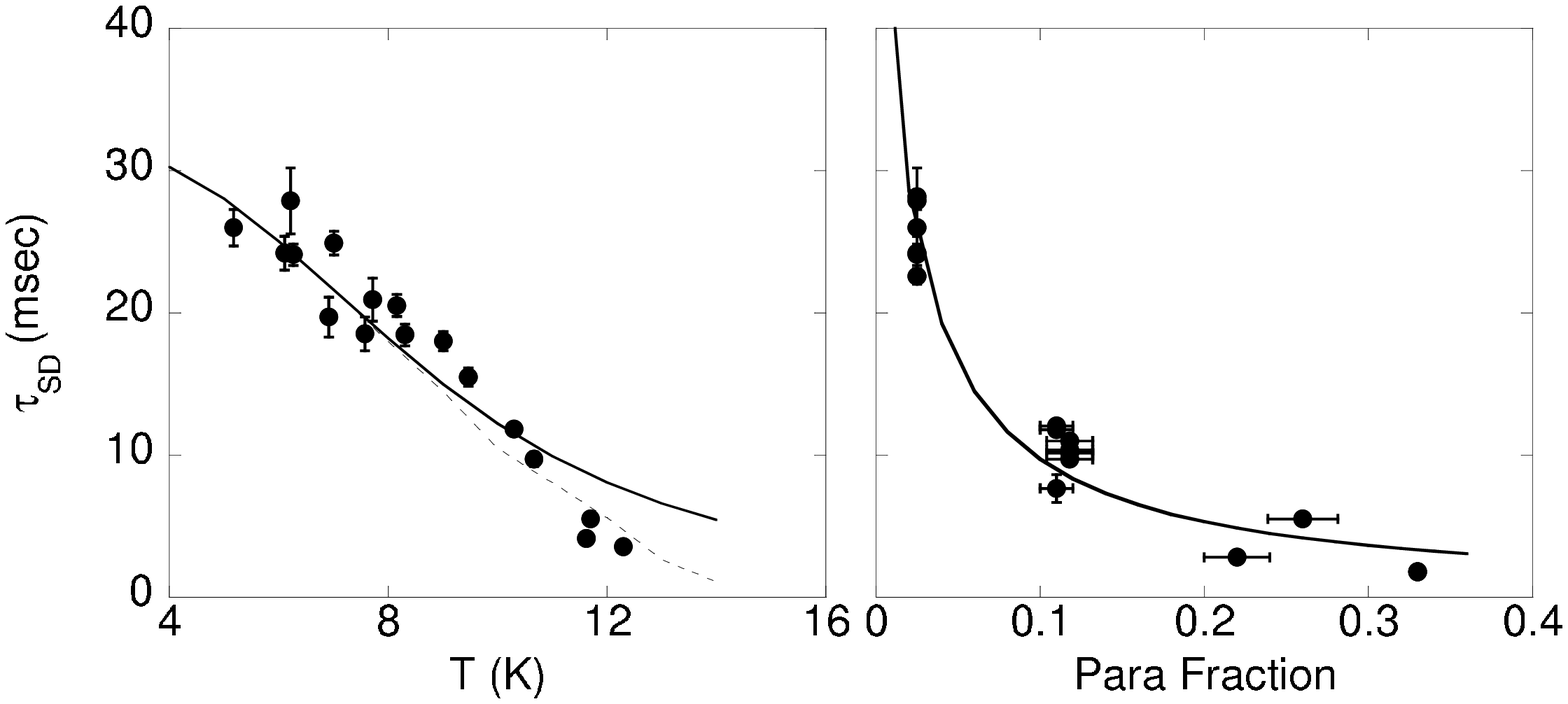}
\caption{\label{SD2LifeTime}Left) Data points are measured SD$_2$ lifetimes as a function of
temperature, with the para-fraction fixed at 2.5\%. Solid lines show the predicted
temperature dependence. Dashed line is
the predicted effect of departure from the solid lifetime model due to
up-scatter from the gas in the guide at higher temperatures.
Right) SD$_2$ lifetimes as a function of para-fraction for all of the data taken below 6K. The solid
line is the model prediction of the para-fraction dependence at an average temperature of
5.6 K }
\end{figure}

Results for UCN lifetimes $\tau _{SD}$ in SD$_2$ as a function
of the SD$_{2}$ temperature and para/ortho-fractions are shown in Fig. \ref
{SD2LifeTime}. The difference between the solid and dashed line demonstrates the
the need to include the effect of deuterium vapor in the guide on the 
lifetime at higher temperatures. With this correction, the
measured lifetimes agree well with theoretical predictions of the up-scatter
rate. The main contributions to the UCN lifetime in SD$_2$ have
been measured and are quantitatively understood. These data demonstrate that
UCN sources based on SD$_2$ converters can provide the promised
large gains over existing sources.

\begin{acknowledgments}
We would like to acknowledge the LANSCE operations staff for
delivering beam in the new pulsed modes needed for this
experiment and Warren Pierce for his excellent machine-shop support. 
We also acknowledge fruitful conversation with R.\ Golub and with A.\ Serebrov. 
This work has been supported by the DOE LDRD program and by the NSF
9420470, 9600202, 9807133, 0071856.
\end{acknowledgments}

\bibliography{ucnlife6}

\end{document}